\documentclass[12pt,preprint]{aastex}

\newcommand{\um}{$\mu$m~}
\newcommand{\ums}{$\mu$m}
\def\kmsMpc{\ifmmode {\rm\,km\,s^{-1}\,Mpc^{-1}}\else
    ${\rm\,km\,s^{-1}\,Mpc^{-1}}$\fi}

\shorttitle{IRS Spectra of Faint IRAS Sources}
\shortauthors{Sargsyan et al.}

\begin{document}

\title{$Spitzer$ IRS Spectra of Optically Faint IRAS Sources}

\author{Lusine Sargsyan\altaffilmark{1}, Areg Mickaelian\altaffilmark{1}, Daniel Weedman\altaffilmark{2}, James Houck\altaffilmark{2}}
\altaffiltext{1}{Byurakan Astrophysical Observatory (BAO) and Isaac Newton Institute (INI) of
Chile, Armenian Branch, Byurakan 378433, Aragatzotn province, Armenia; sarl11@yahoo.com}
\altaffiltext{2}{Astronomy Department, Cornell University, Ithaca, NY 14853; dweedman@astro.cornell.edu}

\begin{abstract}

Extragalactic sources from the IRAS Faint Source Catalog (FSC) which
have the optically faintest magnitudes ($E$ $\ga$ 18) were selected by
spatial coincidence with a source in the FIRST radio survey, and 28
of these sources have been observed with the Infrared Spectrograph
on $Spitzer$ (IRS). While an infrared source is always detected with the IRS at the FIRST
position, only $\sim$ 50\% of the infrared sources are real FSC
detections, as estimated from the number of sources for which the
$f_{\nu}$(25 \ums) determined with the IRS is fainter than the
sensitivity limit for the FSC.  Sources have 0.12 $<$ z $<$ 1.0 and
luminosities (ergs s$^{-1}$) 43.3 $<$ log[$\nu$L$_{\nu}$(5.5 \ums)]
$<$ 46.7, encompassing the range from local ULIRGs to the most
luminous sources discovered by $Spitzer$ at z $\sim$ 2. Detectable PAH features are found in 15 of the sources (54\%), and measurable
silicate absorption is found in 19 sources (68\%); both PAH emission and silicate absorption are present in 11 sources.  PAH luminosities are used to determine the
starburst fraction of bolometric luminosity, and model predictions
for a dusty torus are used to determine the AGN fraction of
luminosity in all sources based on $\nu$L$_{\nu}$(5.5 \ums).
Approximately half of the sources have luminosity dominated by an
AGN and approximately half by a starburst.  The ratio of infrared to
radio flux, defined as q = log[$f_{\nu}$(25 \ums)/$f_{\nu}$(1.4 GHz)],
does not distinguish between AGN and starburst for these sources.

\end{abstract}



\section{Introduction}

As part of efforts to determine the nature of faint infrared
sources, hundreds of spectra have been obtained with the Infrared
Spectrograph on $Spitzer$ (IRS; Houck et al. 2004) of optically faint
sources discovered at 24 $\mu$m with surveys using the Multiband
Imaging Photometer (MIPS; Rieke et al. 2004). The most important result
is that these faint sources having $f_{\nu}$(24 \ums) $\sim$ 1 mJy are generally at high redshift,
typically z $\sim$ 2 \citep{hou05,wee06a,wee06b,yan07,brn07}.
Most show the 9.7\,\um silicate absorption feature, but some have
PAH emission features and some have only a continuum with no
detectable features. These results imply that the sources are
optically faint because they are both distant and obscured by dust.

The absorbed sources and the featureless, power-law sources have been interpreted as obscured AGN \citep{saj07,pol07}. This interpretation implies that the surface density of optically obscured AGN at high redshifts exceeds that for classical, optically discoverable AGN by a factor of two to three in infrared surveys to $f_{\nu}$(24 \ums) $\la$ 1 mJy \citep{wee06c,pol07,fio07}. Extremely luminous starbursts have also been found at z $\sim$ 2 \citep{far08,pop08} which appear similar to local starbursts in all characteristics except luminosity.

To understand these high redshift, dusty sources and to interpret the $Spitzer$-discovered sources within scenarios for evolution in the universe, it is essential to understand how the sources of luminosity are distributed between AGN and starbursts among the overall population of sources. This requires comparison with closer, brighter examples for which AGN and starburst diagnostics within the IRS spectra can be compared to information from other wavelengths.  So far, the comparison samples which show the most similar spectral shapes are the ULIRGS \citep{hao07,ima07,far07,des07}, but the ULIRGs have lower luminosities and smaller redshifts than the sources at z $\sim$ 2.  The ULIRGS observed with the IRS were chosen primarily on the basis of 60 $\mu$m fluxes from the Infrared Astronomical Satellite (IRAS) and were not selected based on optical characteristics.

In this paper, we describe a sample of optically faint,
infrared-bright galaxies which spans luminosities and
redshifts between the ULIRG and high redshift $Spitzer$ IRS samples.
This new sample arises from selecting the optically faintest sources listed in the IRAS Faint Source Catalog
(FSC)\citep{mos90}.  These sources have $E$
$\ga$ 18 mag (the photographic $E$ magnitude band is
centered at 0.66 $\mu$m and approximately corresponds
to $R$ magnitude \citep{hum91}).  

As an additional selection criterion, we also
require that the source be present in the FIRST radio survey
\citep{bec97,whi97}. The requirement of a radio detection was used to
increase confidence that an FSC source is real (as opposed to noise
or cirrus) and to provide an accurate position for selection of the
optical counterpart.

New IRS spectra of 24 sources and 4 spectra of archival
sources are presented. This sample of 28 sources is compared to
previous samples of bright IRAS sources for which $Spitzer$ IRS
spectroscopy is available, and the validity of the FSC
identifications is discussed. Spectral parameters are used to
estimate starburst and AGN contributions to the luminosity.

\section{Sample Selection and Data Analysis}

\subsection{Selection of New Sample of Infrared Bright, Optically Faint Sources}

To select a sample of optically faint infrared sources from the IRAS
FSC, we first cross-correlated the IRAS FSC sources with the FIRST
catalog, using FSC sources listed as real detections rather than upper limits at both 25 $\mu$m
and 60 $\mu$m (sources with quality flags 2 or 3).  2,310 sources
were associated to within 3 times the FSC one sigma positional
uncertainty (ranging from 15\,\arcsec to 90\,\arcsec for individual
sources). Images from the Palomar Second Digitized Sky Survey (DSS2)
were then examined for the 2,310 fields to determine an optical
identification at the radio source position.

Of the 2,310 sources,
1,944 can be identified with bright or medium brightness galaxies
(photographic $E$ $\la$ 16 mag), and 225 are fainter galaxies (16 mag $\ga$ $E$
$\la$ 21 mag).  The remaining 141 sources are identified with bright
Galactic stars or have no optically identifiable
galaxy; the latter could be either spurious sources or cirrus sources. 

The best available optical photometry for the entire FSC
sample is the Minnesota APS catalog \citep{cab03}, which gives
photographic $O$ and $E$ magnitudes. The final sample we select is
defined by having $E$ $>$ 18.  While actual magnitudes
are uncertain to a few-tenths, we can have confidence that the APS
magnitudes are homogeneous and allow location of the optically faintest sources.  The
selection for $E$ $>$ 18 yields 31 sources, which is about 1\% of all sources
common between FSC and FIRST.

\subsection{IRS Observations and Analysis}

We have obtained IRS low-resolution spectra for 24 of these 31 sources, and an additional 4 have been observed in other $Spitzer$ programs.  The sample of objects and the details of the $Spitzer$ IRS\footnote{The IRS was a collaborative venture between Cornell
University and Ball Aerospace Corporation funded by NASA through the
Jet Propulsion Laboratory and the Ames Research Center.} observations are in Table 1. Our new observations were made with the IRS Short Low module in
orders 1 and 2 (SL1 and SL2) and with the Long Low module in orders 1 and 2 (LL1 and
LL2), described in \citet{hou04}.  These give low resolution spectral
coverage from $\sim$5\,\um to $\sim$35\,\um.

Sources were placed on
the slit by using the IRS peakup mode with the red camera.  Background subtraction for spectra was done using coadded background images that added both nod positions having the source in the other slit (i.e., both nods on the LL1 slit when the source is in the LL2 slit produce LL1 spectra of background only).  This subtraction is possible when both SL orders or both LL orders are observed with the same integration times.
Independent extractions of spectra at the two nod positions were compared to reject any highly outlying pixels in either
spectrum, and a final mean spectrum was produced.

Starting with v15 of the SSC basic calibrated data (BCD), spectra were
extracted using the SMART analysis package \citep{hig04}.
Extractions were done with an average width of 4 pixels
(perpendicular to dispersion; width varies with wavelength because
of varying spatial resolution).  Calibrations applied to BCD products use
extractions of 8 pixel width for the calibrating stars, so we
correct fluxes by using Markarian 231 as a standard source and
extracting it with both 8 pixel and 4 pixel windows. Final spectra
were boxcar-smoothed to the approximate resolution of the different
IRS modules (0.2\,\um for SL1 and SL2, 0.3\,\um for LL2, and
0.4\,\um for LL1). All spectra are illustrated in Figures 1-4,
ordered by luminosity.

Measured continuum parameters and feature strengths for the spectra
are in Table 2. Fluxes and equivalent widths (EWs) for the 6.2\,\um and
11.3\,\um PAH features are measured using the single Gaussian line
fit routine within SMART. For the 6.2\,\um PAH line, the fitting is
between 5.5\,\um and 6.9 \um; for the 11.3\,\um line, fitting is between 10.4\,\um and
12.2\,\um.

As discussed by \citet{bra06} and \citet{pop08}, fluxes
and EWs of PAH features depend on the underlying continuum level which is selected.  This is not a straightforward choice and is more
uncertain for the blended 7.7\,\um feature than for the individual
6.2\,\um and 11.3\,\um features.  To minimize uncertainties based on assumptions for fitting the features, we do not list measurements for the
7.7\,\um line.

The fitting procedure we use applies a linear fit for the continuum and a Gaussian fit for the line
between the selected wavelengths. This linear continuum fit is somewhat different from
the spline fit used by \citet{bra06} and yields a higher continuum level. To estimate uncertainties
arising from the fitting procedure, we remeasured the Brandl
starbursts with the procedure we use.  Using the 17 starbursts having
single components (so there is no uncertainty regarding what flux to include in the spectral extraction), the average EW of the 6.2\,\um feature given by
\citet{bra06} is greater by 1.11 then the EWs we measure.

In Table 2, the parameter $\tau_{si}$ describing the depth of the
9.7\,\um silicate absorption feature is measured using the continuum
fitting technique described by \citet{spo07}. The average spectrum
of the 19 sources showing silicate absorption is compared to the
spectrum of Markarian 231 in Figure 5.

We note that Source 9 is a blazar \citep{per78} with observed IRS flux at 25\,\um smaller by a factor of 30 than when observed by IRAS.  The source met the criteria for definition of the sample and so was observed, but this source is not included in the luminosity plots. As a result of the unexpectedly low flux, the IRS spectrum is very noisy and shows no measurable features (Figure 4). PAH limits are not meaningful for this source compared to those for other sources.

\section{Discussion}

\subsection{Reality of IRAS FSC Sources}

The optically faint sources chosen are often near the lower limit of
$f_{\nu}$(25 \ums) in the IRAS FSC; the mean FSC  $f_{\nu}$(25 \ums)
of the sources chosen is 150 mJy, but the typical one sigma noise for an FSC source is
65 mJy. At this low signal to noise, some of these sources may be
spurious. They could appear in the FSC catalog either because of enhancing
noise or because of confusing Galactic cirrus.

To test the reality of the FSC sources, we determined what should be
the IRAS $f_{\nu}$(25 \ums) by applying a synthesized IRAS filter to
the $Spitzer$ IRS spectrum, using the synthesis tool in SMART. The
resulting IRS $f_{\nu}$(25 \ums) are listed in Table 2. In Figure 6, the IRAS $f_{\nu}$(25 \ums) and the synthetic IRS $f_{\nu}$(25 \ums) are compared.

Fig. 6 shows that the $f_{\nu}$(25 \ums) from IRAS are always greater than the $f_{\nu}$(25 \ums) measured with the $Spitzer$ IRS, with a median factor of 1.6 and extending to a factor of 6. The discrepancy increases with decreasing flux, so that the faintest sources show the largest ratio of IRAS to $Spitzer$ flux.  Various previous observations with comparable IRS signal to noise indicate internal consistency of $Spitzer$ MIPS and IRS fluxes at 24\,\um to within 5\% \citep{hou07}, so IRS flux calibration uncertainty is much less than the differences observed with IRAS fluxes.

One possible source of difference arises when the IRAS source is larger than the IRS slit so that so that some flux is omitted in the IRS spectrum. This is the case for many nearby starbursts in \citet{bra06}. Images of the FSC sources were obtained at $\sim$ 25\,\um because acquisition of sources used the IRS red peakup camera, and all images show an unresolved source as evidenced by the presence of a diffraction ring.  This indicates that extended sources cannot explain the differences between IRAS and IRS fluxes.

We conclude that the differences between IRAS and IRS fluxes arise because many of these sources are near the detection limit for the IRAS FSC.  Fig. 6 shows that 14 of the IRAS FSC sources are actually fainter at 25  \ums, as measured by IRS $f_{\nu}$(25 \ums), than the typical one sigma sensitivity limit of 65 mJy quoted for the FSC.  This implies that these faint FSC "detections" are spurious, perhaps accidental noise spikes.

Therefore, we use the ratio IRAS $f_{\nu}$(25 \ums)/IRS $f_{\nu}$(25 \ums) to estimate the reality of the IRAS FSC source.  If this ratio is $<$ 1.5, we assume the source is really detected in the IRAS FSC, because flux discrepancies of this amount are reasonable for the faint FSC sources.  Also, this is the approximate flux ratio at which the IRAS FSC fluxes fall below the 65 mJy sensitivity limit in Figure 6.  By this criterion, 14 (50\%) of the sources are real FSC detections (counting the blazar for which the flux discrepancy is attributed to variability).

If half of the FSC infrared sources are not real, why did all of the FSC sources appear as $Spitzer$ sources?  We believe that the answer is because the requirement of a FIRST radio source at the IRAS FSC position was also a criterion for source selection.  As shown below, $Spitzer$ sensitivities are such that a $Spitzer$ source should be detected for any radio source brighter than the FIRST limit, regardless of whether the source was actually detected in the IRAS FSC.

The ratio of infrared to radio flux is described by the parameter q
= log[$f_{\nu}$(25 \ums)/$f_{\nu}$(1.4 GHz)].  The median q = 0.8 for
faint $Spitzer$  sources detected at both 24\,\um and 1.4 GHz
\citep{app04}, and this is also the q value expected from previously
known radio-infrared correlations for starbursts \citep{con82}.
The FIRST catalogue includes sources to a point-source limit of 1 mJy for typical
one sigma rms of 0.13 mJy \citep{whi97}. This limit means that a
$Spitzer$ source would have been found with $f_{\nu}$(25 \ums) $>$ 6
mJy for any FIRST source with q $>$ 0.8. Such a source would be
detectable by the peakup pointing of $Spitzer$, and an IRS spectrum would be
obtained, even if the FSC detection were not real.

This result means that any FIRST source with an infrared to radio
ratio exceeding the median expected value would be detected in our
$Spitzer$ pointings. The existence of a FIRST source makes it
probable, therefore, that a detectable $Spitzer$ source would be
found even if the FSC source used for the initial identification
were not real. This explains why $Spitzer$ sources were always found
at the FIRST position even in those cases when the $f_{\nu}$(25
\ums) are too faint for the FSC detection to be real.

\subsection{Characteristics of Spectra and Near-Infrared Luminosities}

Spectra of the FSC sample are shown in Figures 1 through 4, ordered by $\nu$L$_{\nu}$(5.5 \ums), with lowest luminosity sources shown first. This ordering allows some overall spectral trends to be seen.  Sources with the strongest PAH features and weakest silicate absorption (sources 15, 5, and 7) are among the lowest luminosity sources, in Figure 1.  Sources of intermediate luminosity, in Figures 2 and 3, generally show conspicuous silicate absorption and weak PAH features.  Finally, the highest luminosity sources (Figure 4) have the weakest silicate absorption and the weakest PAH features.

As a comparison standard throughout, we use the thoroughly studied ULIRG Markarian 231 because its mid infrared spectrum \citep{wee05,arm07} is similar in shape to many other absorbed sources.  In Figure 5, we show this similarity by comparing the average spectrum of all the FSC sources in Table 2 having silicate absorption to the spectrum of Markarian 231.  The average absorption depth is similar to that of Markarian 231, for which the silicate optical depth of -0.7 corresponds to maximum extinction by the silicate feature of $\sim$ 50\% of the continuum flux.

The distribution of continuum luminosities for the FSC sources is shown in Figure 7, based on $\nu$L$_{\nu}$(5.5 \ums).  This parameter is chosen for comparison to the large samples of ULIRGs, Sy1, Sy2, and QSOs summarized in \citet{hao07}. All of these sources previously observed with the IRS have log[$\nu$L$_{\nu}$(5.5 \ums)] $<$ 46.1, and the median ULIRG luminosity is log[$\nu$L$_{\nu}$(5.5 \ums)] = 44.4.  The luminosity limit of these previously observed sources is exceeded by 3 of the FSC sources.  The most luminous obscured sources at z $\sim$ 2 \citep{hou05,yan07} have log[$\nu$L$_{\nu}$(5.5 \ums)] $<$ 46.6 \citep{pol07}, and this extreme luminosity is matched by one of the FSC sources.  The median luminosity of the FSC sources in Figure 7 is log[$\nu$L$_{\nu}$(5.5 \ums)] = 44.8. These results indicate, therefore, that the FSC sample encompasses a luminosity range which includes typical ULIRGs and extends to the most luminous sources discovered by $Spitzer$.

\subsection{Comparison of FSC Sample to Previous ULIRG samples}

One objective of the present study was to assemble a sample of IRAS sources substantially fainter in infrared and optical flux than the luminous ULIRGS  defined by the IRAS 1 Jy survey \citep{kim98} whose IRS spectra are discussed by \citet{ima07}, \citet{far07}, and \citet{des07}.  Objectives include learning how the nature of IRAS sources changes as flux becomes fainter, and especially to seek sources that are more similar in luminosity and redshift to the high-redshift sources discovered in $Spitzer$ surveys of sources at $\sim$ 1 mJy.

The FSC sample was chosen to have the faintest optical magnitudes that could be found in the FSC in order to select the IRAS sources with the most extreme values of observed infrared to optical flux ratio (IR/opt). The faint $Spitzer$ samples in Bootes \citep{hou05} and the FLS \citep{yan07} have typical $R$ $\sim$ 25 mag and $f_{\nu}$(25 \ums) $\sim$ 1 mJy.  In the observer's frame, objects having similar IR/opt in the FSC, with typical $f_{\nu}$(25 \ums) $\sim$ 100 mJy, should have $R$ $>$ 20 mag. Very few FSC objects are so faint and so extreme in IR/opt, but all sources in Table 1 have photographic $E$ $>$ 18. (We assume within the magnitude uncertainties that $E$ and $R$ are similar).

Figure 8 compares the FSC sample and the ULIRG samples in 25\,\um
flux and redshift (using the $Spitzer$ IRS $f_{\nu}$(25 \ums) for
the FSC sources and the IRAS $f_{\nu}$(25 \ums) for the ULIRG
sources).  The FSC sample extends to higher redshifts than the
previously published ULIRG samples which have IRS spectra.  The maximum redshift
of the FSC sample is 0.93 with median redshift of 0.25 whereas the
ULIRG samples tabulated by \citet{ima07} and \citet{far07} have
maximum z of 0.26. The combined dataset extends over a range of
nearly 1000 in $f_{\nu}$(25 \ums).

The ratio of mid-infrared to optical or near-infrared flux is a measure of how much a source is dominated by dust absorption and dust continuum luminosity.  In general, very dusty sources will have ratios enhanced both by increased extinction of shorter wavelengths and increased dust continuum at longer wavelengths, although detailed understanding of individual sources requires modeling of the dust absorption and emission contributions \citep[e.g. ][]{mar07}. For comparing fluxes at different wavelengths among samples, homogeneous all-sky data measuring optical magnitudes for IRAS sources are not available other than the photographic magnitudes used for our initial selection.  However, the 2MASS survey \citep{skr06} encompasses all sources in the ULIRG samples and most sources in our FSC sample with near-infrared J,H,K photometry.

In Figure 9, the flux distributions at 25\,\um and ratio  $f_{\nu}$(25 \ums)/$f_{\nu}$(J) for the FSC sample are compared with the ULIRG samples in \citet{ima07} and \citet{far07}. (The flux transformation adopted for 2MASS J mag is that zero magnitude corresponds to 1590 Jy.)
The median for the combined ULIRG samples is $f_{\nu}$(25 \ums)/$f_{\nu}$(J) = 2.42 and is the same for the FSC sample, counting limits.  Although the FSC sample was chosen for faint optical magnitudes, this similarity in $f_{\nu}$(25 \ums)/$f_{\nu}$(J) indicates that the FSC sample does not select in favor of sources with a greater dust content compared to the ULIRG samples.  The FSC sample does contain sources at greater redshifts than the \citet{ima07} and \citet{far07} ULIRG samples, as shown in Fig. 8, although this result arises in part because the \citet{ima07} sample was restricted to ULIRGs with z $<$ 0.15.

\subsection{Luminosities from AGN and Starbursts}

One major objective of previous ULIRG studies has been to determine the relative contributions of starbursts and AGN to the total luminosity of a source.  A division of luminosity between starbursts and AGN at rest frame $\sim$ 6\,\um can be determined by the strength of the PAH features compared to the continuum, as measured by the equivalent width (EW) of the PAH features \citep{gen98,lau00,ima07,far07,des07}.  Pure starbursts with no indicators of AGN at any wavelength have rest frame EW(6.2 \ums) $>$ 0.45 \citep{bra06}. Smaller equivalent widths correspond to an increasing continuum, assumed to be caused by an AGN.  Very few ULIRGS have such strong PAH features; only 2 of 49 in \citet{ima07} and 8 of 107 in \citet{des07}, which indicates a substantial AGN contribution to the continuum at 6\,\um.

The distribution of PAH strength for the FSC sample is shown in Figure 7; only one of 28 sources has 6.2\,\um EW $>$ 0.45 and only three have EW $>$ 0.2.  These results indicate that, as in the ULIRG samples, the near-infrared continuum is generally much stronger relative to the PAH feature than in pure starbursts.   It is also seen in Figure 7 that the most luminous sources at rest frame 5.5\,\um in the FSC sample are sources with the smallest PAH equivalent widths. The simplest interpretation of these results is that the most luminous near-infrared sources are dominated by AGN power.

How the luminosity divides in the near-infrared between AGN and starbursts is not necessarily the same as the separation of starburst and AGN power for bolometric luminosities. Many efforts have been made to separate the bolometric contributions within ULIRGs of AGN and starbursts by assuming templates for the total AGN spectrum and the total starburst spectrum \citep[e.g.][]{far03}.  In general, results are that the far infrared luminosity from cooler dust is attributed primarily to the starburst whereas the near infrared luminosity from warmer dust is attributed to an AGN. The warmer AGN contribution leads to stronger near infrared continuum than from starbursts, causing the low equivalent widths for PAH features.

Assumed templates can be incorrect if, for example, there is cooler dust outside of the dusty torus which is also heated by the AGN, or if starbursts are deeply buried in optically thick clouds that produce hotter dust \citep[e.g.][]{lev07,ima07,pol07}. The templates may be examples of observed sources, but the intrinsic nature of the template sources may not be fully understood.  It is desirable, therefore, to determine an estimate of AGN and starburst contributions which is independent of template assumptions.

For the rest frame parameters $\nu$L$_{\nu}$(5.5 \ums) and L(6.2 \ums), which are measured for our sources in Table 3, empirical determinations relate these parameters to the bolometric luminosities of obscured AGN [L$_{ir}$(AGN)] and of starbursts [L$_{ir}$(SB)].  For local starbursts \citep{bra06} and high redshift submillimeter galaxies which are starbursts \citep{pop08}, an empirical relation has been found to be log[L$_{ir}$(SB)] = log[L(6.2 \ums)]+ 2.7$\pm$0.1; L(6.2 \ums) is the total luminosity of the 6.2\,\um PAH feature, after subtracting the underlying continuum.

This result derives from an observational comparison of PAH luminosity and bolometric luminosity; while there may be some extinction of
the 6.2\,\um feature, effects of extinction would be included in the empirical result.  Adopting this relation also assumes
that L(6.2 \ums) is measured the same in all sources.  The uncertainty listed does not include the $\sim$ 10\% systematic uncertainty in measurement of 6.2\,\um fluxes that may arise from differences in the fitting procedure discused in section 2.2.  \citet{pop08}
state that their measures are "consistent" with those of \citet{bra06}, to which we also compare in section 2.2.  Because our measures agree within 10\% of the Brandl measures, we adopt the relation in \citet{pop08} without applying a systematic
correction.  This relation applies to a "pure" starburst, with no AGN
contribution to infrared luminosity.

For AGN, it is assumed that the infrared continuum associated with the AGN arises from warm dust within a torus surrounding the accretion region.  Using the clumpy torus model of \citet{hon06}, empirical fits to overall SEDs in \citet{pol07} give log[L$_{ir}$(AGN)] = log[$\nu$L$_{\nu}$(6.0 \ums)] + 0.32$\pm$0.06.  We make a nominal modification to this by transforming $\nu$L$_{\nu}$(5.5 \ums) to $\nu$L$_{\nu}$(6.0 \ums) using Markarian 231 because of its similarity to the average spectrum of our absorbed sources in Figure 10.  With this correction, we have log[L$_{ir}$(AGN)] = log[$\nu$L$_{\nu}$(5.5 \ums)] + 0.33$\pm$0.06.  We adopt this relation as applying to a "pure" AGN, with no contribution to infrared luminosity from any source other than the torus heated by the AGN.

These two relations for deriving L$_{ir}$(SB) and L$_{ir}$(AGN) are used to determine the starburst and AGN luminosities for the FSC sources. Results are in Table 3 and Figure 10.  We find that sources divide equally between those dominated by AGN luminosity and those by starburst luminosity.  It is notable in Figure 10 that the median luminosity in the sample attributed to AGN, log L$_{ir}$(AGN) $\sim$ 45.2, is the same as that attributed to starbursts, log L$_{ir}$(SB), although AGN sources reach to higher luminosities.





\subsection{Radio Characteristics}

Because this FSC sample was selected with the requirement of a FIRST radio detection \citep{whi97}, all sources have 1.4 GHz detections.  We list in Table 2 the values of the parameter q, defined as q = log[$f_{\nu}$(25 \ums)/$f_{\nu}$(1.4 GHz)] in the observed frame.  For determining q, sources are assumed to be unresolved so the peak $f_{\nu}$(1.4 GHz) is used.  For the FSC sources, the peak flux typically differs by $\sim$ 10\% from integrated fluxes. As discussed in section 3.1, the much higher S/N and higher spatial resolution of the $Spitzer$ detections gives more accurate photometry at 25 $\mu$m than the IRAS FSC fluxes so the IRS $f_{\nu}$(25 \ums) are used to determine q.

The comparison of q and $\nu$L$_{\nu}$(5.5 \ums) is shown in Figure 11.  Larger values of q correspond to relatively weaker radio sources. The median value of q for the FSC sample is 1.25.  This is larger than the median of q = 0.8 for faint $Spitzer$ First Look Survey sources detected at both 24\,\um and 1.4 GHz \citep{app04}, which is the q value expected from previously known radio-infrared correlations for starbursts \citep{con82}.  The median q is even less, q = 0.4, for faint sources in the $Spitzer$ First Look Survey which have IRS spectra and z $>$ 1 in \citet{wee06c}.  Some of these differences may be redshift effects.  For example, Markarian 231 (shown in Figure 9) has q = 1.6 but, if at z = 2, would have q = 0.6.

The results in Fig. 11 do not indicate that the value of q can be used as a discriminant between starburst and AGN sources. A median q of $\sim$ 1.3 is found both for the starburst sources that have PAH features and for the remaining sources dominated by AGN.  Also, the range of q is similar for both categories of sources.  All of the values of q for the FSC sources are much larger than in radio-loud AGN or quasars, for which q $<$ -1 at any redshift \citep{hig05}. The radio data, therefore, do not indicate any evidence of radio-loud AGN among the FSC sources.

\section{Summary}

The IRAS Faint Source Catalog (FSC) was used to select the 31 optically faintest FSC sources which are also identified in the FIRST radio survey, representing the optically faintest 1\% of IRAS extragalactic sources.  28 of these sources have been observed with the Infrared Spectrograph on $Spitzer$. $Spitzer$ fluxes indicate that about half of the sources have $f_{\nu}$(25 \ums) which are below the IRAS FSC flux limit.  This indicates that about half of the optically faint sources taken from the FSC are actually spurious FSC detections.  These sources are nevertheless detected by $Spitzer$ because the presence of a FIRST radio source also provides an infrared source of sufficient brightness for a $Spitzer$ detection.

This FSC sample reaches to higher redshifts and higher luminosities than the IRAS-discovered ULIRG samples
observed spectroscopically with $Spitzer$, and overlaps the
luminosity range of $Spitzer$-discovered 24\,\um sources at z $\sim$ 2. The
FSC sources have 0.12 $<$ z $<$ 1.0 and luminosities 43.3 $<$
log[$\nu$L$_{\nu}$(5.5 \ums)] $<$ 46.7.  15 of the sources have
detectable PAH features, and 19 have measurable silicate absorption.
Median properties of the sample having silicate absorption are very
similar to the ULIRG Markarian 231 in silicate strength, continuum
luminosity, ratio $f_{\nu}$(25 \ums)/$f_{\nu}$(J), and relative
radio luminosity.

PAH luminosities are used to determine the starburst luminosity within each source, and predictions from dusty torus models are used to determine the AGN  luminosity.  Sources have similar bolometric luminosities arising from starbursts and from AGN and are equally divided between sources dominated by starbursts and sources dominated by AGN.  The ratio of infrared to radio flux is not a measure of whether sources are dominated in the infrared by starburst or AGN luminosity.

\acknowledgments
We thank Don Barry for technical assistance. This work is based on observations made with the
Spitzer Space Telescope, which is operated by the Jet Propulsion
Laboratory, California Institute of Technology under NASA contract
1407.  Support for this work by the IRS GTO team at Cornell University was provided by NASA through Contract
Number 1257184 issued by JPL/Caltech. Support was also provided by the US Civilian Research and Development Foundation under grant ARP1-2849-YE-06. This research has made use of the NASA/IPAC Extragalactic Database (NED) which is operated by the Jet Propulsion Laboratory, California Institute of Technology, under contract with the National Aeronautics and Space Administration.

\clearpage

\begin{deluxetable}{lcccccc} 
\tablecolumns{7}
\tabletypesize{\footnotesize}
\tablewidth{0pc}
\tablecaption{IRS Observations of Sources}
\tablehead{
\colhead{Number}& \colhead{Source Name\tablenotemark{a}}& \colhead{coordinates\tablenotemark{b}} & \colhead{AOR}& \colhead{program}& \colhead{time\tablenotemark{c}} & \colhead{date\tablenotemark{d}}\\
  \colhead{}& \colhead{}& \colhead{FIRST J2000} &\colhead{} & \colhead{}& \colhead{s} & \colhead{mo/day/yr}
}
\startdata
1 & FSC 07247+6124 & 072912.10+611853.5 & 17540096 & 30121 & 120,60 & 11/16/06\\
2 & FSC 09105+4108 & 091345.28+405632.6 & 7771392 & 1018 &120,180 &11/29/03 \\ 
3 & FSC 09121+2430 & 091501.71+241812.2 &17540352 & 30121 & 84,60 & 05/01/07\\
4 & FSC 09235+5425 & 092703.07+541206.6  &17540608 & 30121 & 120,60 & 11/17/06\\
5 & FSC 09284+0413 & 093101.27+035955.2 &17540864 & 30121 & 84,60 & 06/09/07\\
6 & FSC 09425+1751 & 094521.36+173753.4  &17541120 & 30121 & 84,60 & 06/08/07\\
7 & FSC 10219+2657& 102447.39+264209.0 &17541888 & 30121 & 84,60 & 12/25/06\\
8 & FSC 11257+5113 & 112832.73+505721.1 &17542400 & 30121 & 120,60 & 05/02/07\\
9 & FSC 13080+3237 & 131028.64+322049.2 &17542912 & 30121 & 120,60& 06/27/06\\
10 & FSC 13297+4907 & 133150.54+485150.6 &17543168 & 30121 & 120,60 &06/27/06\\
11 & FSC 14448-0141 & 144727.54-015330.3 &17543424 & 30121 & 120,60 & 07/25/06\\
12 & FSC 14475+1418 & 144954.86+140610.5&17543680 & 30121 & 120,60 & 07/29/06\\
13 & FSC 14481+4454 & 144953.70+444150.3 & 17543936 & 30121 & 120,60 & 07/04/06\\
14 & FSC 14503+6006& 145135.04+595437.6 &  17544192 & 30121 & 120,60 & 07/04/06\\
15 & FSC 14516+3851& 145335.96+383913.1 & 17544448 & 30121 & 120,60 & 06/24/06\\
16 & FSC 14589+2329 & 150113.19+232908.2 & 4169216 & 49 & 56,120 & 02/07/04\\ 
17 & FSC 15065+3852 & 150825.42+384122.1 & 17544704  & 30121 & 120,60 & 07/01/06\\
18 & FSC 15307+3252 & 153244.05+324246.7 & 4983552 & 105 & 360,480 &03/04/04\\
19 & FSC 15385+4320 & 154014.02+431042.4 &  17544960 & 30121 & 120,60 & 06/25/06\\
20 & FSC 15458+0041 & 154823.38+003212.8 & 17545216 & 30121 & 120,60 & 03/20/07\\
21 & FSC 15492+3454 & 155108.86+344533.6 & 17545472 & 30121 & 120,60 & 06/25/06\\
22 & FSC 15496+0331 & 155206.16+032244.0 & 17545728 & 30121 & 120,60 & 03/16/07\\
23 & FSC 15585+4518 & 160003.29+451046.2 & 17545984 & 30121 & 120,60 & 07/24/06\\
24 & FSC 16001+1652 & 160222.38+164354.3 & 17546240 & 30121 & 120,60 & 03/10/07\\
25 & FSC 16073+0209 & 160949.75+020130.8 & 17546496 & 30121 & 120,60 & 03/18/07\\
26 & FSC 16156+0146 & 161809.36+013922.1 & 17546752 & 30121 & 120,60 & 09/17/06\\ 
27 & FSC 16242+2218 & 162626.00+221145.9 & 17547008 & 30121 & 120,60 & 09/17/06\\
28 & FSC 17233+3712 & 172507.40+370932.1 & 4986880 & 105 & 360,480 &06/07/04 \\

\enddata
\tablenotetext{a}{Source name from the IRAS Faint Source Catalog.}
\tablenotetext{b}{Coordinates for the FIRST source which is closest to the FSC source.}
\tablenotetext{c}{Total integration time for IRS short low modules (first number) and long low modules (second number).}
\tablenotetext{d}{Date of IRS observation}
\end{deluxetable}

\clearpage

\begin{deluxetable}{lcccccccccccccc} 
\rotate
\tablecolumns{15}
\tabletypesize{\tiny}
\tablewidth{0pc}
\tablecaption{Observed Properties of Sources}
\tablehead{
\colhead{No.}& \colhead{Source}& \colhead{z} & \colhead{$f_{\nu}$(25 \ums)} & \colhead{$f_{\nu}$(25 \ums)} & \colhead{$E$\tablenotemark{a}}&  \colhead{$J$\tablenotemark{b}} &\colhead{$f_{\nu}$(1.4 GHz)} &\colhead{q\tablenotemark{c}} &  \colhead{$f_{\nu}$(5.5 \ums)} & \colhead{$f_{\nu}$(15 \ums)} &\colhead{EW(6.2\ums)}  & \colhead{f(6.2\ums)\tablenotemark{d}} & \colhead{f(11.3\ums)\tablenotemark{d}} & \colhead{$\tau_{si}$\tablenotemark{e}} \\
  \colhead{}& \colhead{}& \colhead {} & \colhead{IRAS mJy} &\colhead{IRS mJy} & \colhead{mag}& \colhead{mag} &\colhead{mJy} &\colhead{} & \colhead{mJy} & \colhead{mJy}&\colhead{\ums} & \colhead{} & \colhead{} & \colhead{}
}
\startdata
1 & 07247 & 0.137 & 175 & 119 &18.0 & 16.08 &3.13 &1.57 &4.1 & 47.2 & 0.08$\pm$0.01 &2.94$\pm$0.42 & 2.48$\pm$0.46 &-0.31 \\
2 & 09105&  0.446 & 333 & 333 & 18.1 & 16.83 &5.79 &1.76 &54 & 297 & $<$ 0.002 &$<$ 1.3 & $<$ 0.2 &-0.34 \\
3 & 09121 & 0.84\tablenotemark{f} & 179 & 100 &20.0 & 16.56 &9.80 &1.01 & 42.5 & 119 & $<$ 0.01 & $<$ 1.8 & $<$ 0.4 & 0.0 \\
4 & 09235 &  0.123 & 136 & 97 &18.3 & 16.75 & 3.31&1.47 & 2.0 & 42.8 & 0.19$\pm$0.02 &4.62$\pm$0.47 & 2.67$\pm$0.16 & 0.0 \\
5 &  09284 &  0.146 & 147 & 25 &18.2 & 16.11 &1.05 &1.38 & 1.5 & 6.9 & 0.56$\pm$0.03 &12.14$\pm$0.73 & 10.0$\pm$0.2 & 0.0 \\
6 &  09425  & 0.130 & 436 & 322 &18.6 & 15.72 &38.7 &0.92 & 17.8 & 162 & $<$ 0.1 & $<$ 0.3 & $<$ 3.91 &-0.78\\
7 & 10219 & 0.225 & 151 & 28 &18.8 & 16.77 &1.68 &1.22 & 1.3 & 11.8 & 0.29$\pm$0.04 &4.53$\pm$0.7 & 3.92$\pm$0.5 & 0.0\\
8 &  11257  & 0.197 & 123 & 75 &18.4 & 16.55 & 1.77&1.63 & 3.0 & 35.8 & 0.07$\pm$0.01 &3.12$\pm$0.6 & 1.91$\pm$0.4 &-0.51 \\
9 &  13080 & 0.997\tablenotemark{f} & 271 & 9.0 &19.2 & 16.34 & \nodata &\nodata &3.5 & 10.3 & \nodata &\nodata & \nodata & 0.0 \\ 
10 &  13297 & 0.128 & 139 & 102 &18.5 & 16.68 &1.17 &1.94 & 3.6 & 53.1 & $<$ 0.05 & $<$ 1.8 & $<$ 1.5 &-0.69 \\
11 &  14448 & 0.210 & 176 & 38 &18.9 & $>$ 17.7 &1.78 &1.33 & 3.8 & 23.4 & $<$ 0.02 & $<$0.9 & 2.01$\pm$0.65 &-0.62 \\
12 &  14475 & 0.251 & 143 & 114 &18.5 & 16.68 &12.8 &0.95 & 11.6 & 91 & 0.03$\pm$0.005 &3.4$\pm$0.52 & $<$ 3.2 &-0.89 \\
13 &  14481 &  0.670 & 85 & 64 &21.0 & $>$ 17.7 & 10.7 & 0.78 & 19.2 & 67.6 & $<$ 0.006 & $<$ 1.1 & $<$ 0.5 & 0.0 \\
14 &  14503 &  0.577 & 79 & 58 &21.0 & $>$ 17.7 & 16.3 &0.55 & 8.4 & 60 & $<$ 0.02 &  $<$ 1.9 & $<$ 1.0 & -1.01 \\
15 &  14516  & 0.153 & 85 & 24 &19.3 & 16.77 &2.07 & 1.06& 0.75 & 9.4 & 0.20$\pm$0.03 &2.24$\pm$0.38 & 2.49$\pm$0.14 &-0.87 \\
16 &  14589 & 0.261 & 84 & 53 &17.8 & 15.98 & 4.49&1.07 &11.7 & 37.8 & $<$ 0.017 & $<$ 1.73 & $<$ 3.4 &-0.24 \\
17 &  15065  & 0.355 & 82 & 45 &20.2 & $>$ 17.7 &2.43 &1.27 & 4.9 & 32.9 &$<$ 0.013 & $<$ 0.84 & 4.41$\pm$0.27 &-0.59 \\
18 &  15307  & 0.927 & 71 & 46 & 19.0 & $>$ 17.7 & 5.7&0.91 &9.4 & 70.0 &  $<$ 0.006 & $<$ 0.5 & $<$ 2.4 &-0.32 \\
19 &  15385  & 0.380 & 92 & 76 &18.5 & $>$ 17.7 &1.10 &1.84 & 6.0 & 54.6 & 0.02$\pm$0.005 & 1.22$\pm$0.34 & $<$ 1.1 & 0.0 \\
20 &  15458  & 0.254 & 126 & 40 &18.7 & 16.70 &2.53 &1.20 & 4.3 & 25.2 & $<$ 0.01 & $<$ 0.59 & $<$ 0.26 & 0.0 \\
21 &  15492 & 0.311 & 80 & 40 &19.6 & $>$ 17.7  &1.08 &1.57& 6.0 & 29.7 & $<$ 0.02 &$<$ 0.9 & 1.56$\pm$0.42 &-1.46 \\
22 & 15496 & 0.193 & 144 & 122 &18.5 & 16.20 &28.5 & 0.63 & 13.7 & 86.7 & $<$ 0.01 &$<$ 1.8 & 2.39$\pm$0.47 & -1.10 \\
23 & 15585 &  0.486 & 64 & 11.0 &20.0 & $>$ 17.7 &1.58 & 0.87 & 3.1 & 11.9 & 0.10$\pm$0.04 &2.0$\pm$0.82 & 3.25$\pm$0.51 & -2.2 \\
24 &  16001 &  0.672 & 144 & 104 &18.1 & 16.62 &1.90 &1.74 & 22.9 & 109 & $<$ 0.01 & $<$ 0.25 & $<$ 1.29 & 0.0 \\
25 & 16073 & 0.223 & 153 & 123  &19.2 & 17.46 &5.41 &1.36 &9.6 & 96 & 0.06$\pm$0.002 & 7.5$\pm$0.3 & $<$ 4.1 & -1.5 \\
26 &  16156 & 0.133 & 279 & 261  &18.3 & 16.57 &7.9 &1.52 & 27 & 154 & $<$ 0.01 & $<$ 2.8 & $<$ 0.8 & -2.6 \\
27 &16242  & 0.157 & 71 & 31 &19.5 & $>$ 17.7 &1.39 &1.35 &1.3 & 15.1 & $<$ 0.01 & $<$ 0.25 & 1.07$\pm$0.12 & -2.7 \\
28 &  17233  & 0.702 & 50 & 29 & 21.0 & $>$ 17.7 &3.16 &0.96 &1.7 & 34.3 & $<$ 0.04 & $<$ 0.78 & $<$ 1.2 &-0.51 \\
\enddata
\tablenotetext{a}{Photographic $E$ magnitude from APS.}
\tablenotetext{b}{J magnitude from 2MASS (Skrutskie et al. 2006).}
\tablenotetext{c}{q = log[$f_{\nu}$(25 \ums)/$f_{\nu}$(1.4 GHz)] in observed frame, using $f_{\nu}$(1.4 GHz) of "peak flux" from FIRST, corresponding to an unresolved source.}
\tablenotetext{d}{Total flux of feature in units of 10$^{-21}$W cm$^{-2}$, fit with single Gaussian. Uncertainties in fluxes and equivalent widths are statistical uncertainties from fitting of feature and do not include systematic uncertainties of $\la$ 10\% which may arise from fitting procedure chosen, as discussed in text.}
\tablenotetext{e}{$\tau_{si}$ is measure of depth of 9.7\,\um absorption feature, defined as $\tau_{si}$ = ln[$f_{\nu}$(abs)/$f_{\nu}$(cont)], for $f_{\nu}$(abs) measured at the maximum depth of absorption and $f_{\nu}$(cont) the unabsorbed continuum as extrapolated in \citet{spo07}.}
\tablenotetext{f}{Redshift for no. 3, a red quasar, is optical redshift from \citet{urr05}, and for number 9, a blazar, is from \citet{per78}; other redshifts are determined from the IRS spectra.}

\end{deluxetable}

\clearpage

\begin{deluxetable}{lccccccccc} 
\rotate
\tablecolumns{10}
\tabletypesize{\tiny}
\tablewidth{0pc}
\tablecaption{Luminosities of Sources}
\tablehead{
\colhead{Number}& \colhead{Source}& \colhead{D$_L$\tablenotemark{a}} & \colhead{log[$\nu$L$_{\nu}$(5.5 \ums)]} & \colhead{log[$\nu$L$_{\nu}$(15 \ums)]} & \colhead{log[L(6.2)]} & \colhead{log[L(6.2+11.3)]}&  \colhead{log[L(SB)]\tablenotemark{b}} & \colhead{log[L(AGN)]\tablenotemark{c}} & \colhead{L(SB)/[L(SB)+L(AGN)]\tablenotemark{d}}  \\
  \colhead{}& \colhead{}& \colhead {Mpc} & \colhead{ergs s$^{-1}$} & \colhead{ergs s$^{-1}$} &\colhead{ergs s$^{-1}$} & \colhead{ergs s$^{-1}$}& \colhead{ergs s$^{-1}$} &\colhead{ergs s$^{-1}$} &\colhead{}
}
\startdata
1 &  07247 & 637 & 43.98 & 44.60 &42.15 & 42.42 & 44.85 & 44.31 & $>$ 0.78 \\ 
2 &  09105 & 2460 & 46.17 & 46.47 &$<$42.96 & $<$43.03 & $<$45.66 & 46.50 &$<$ 0.13\\ 
3 &  09121 & 5340 & 46.63 & 46.64 &$<$43.79 & $<$43.88 & $<$46.49 & 46.96 & $<$ 0.25\\ 
4 &  09235 & 565 & 43.58 & 44.46 &42.25 & 42.44 &44.95 & 43.91 & $>$ 0.92\\
5 &  09284 &683  & 43.60 & 43.83 & 42.83& 43.09 &45.53 & 43.93 & $>$ 0.98\\
6 &  09425 & 602 & 44.57 &45.09  &$<$41.10 & $<$42.26 &$<$43.80 & 44.90 & $<$ 0.07\\
7 &  10219  & 1108 & 43.93 &44.45  &42.82  & 43.09 & 45.52 & 44.26 & $>$ 0.95\\
8 &  11257 &952  & 44.17 &44.81  &42.53 & 42.73 &45.23 & 44.50 & $>$ 0.84\\
9 &  13080 & 6610 & 45.70 & 45.73 & \nodata& \nodata &\nodata &\nodata &\nodata \\
10 &  13297  & 594 &43.87  &44.60  & $<$41.87& $<$42.14 &$<$44.57 & 44.20 &$<$ 0.70\\
11 &  14448 & 1025 & 44.34 &44.68  &$<$42.05 &$<$42.56 & $<$44.75 & 44.67 & $<$ 0.55\\
12 &  14475 &1254  & 44.98 &45.43  & 42.80& $<$43.09 & 45.50 & 45.31 & $>$ 0.61 \\
13 &  14481  &4036  &46.08  &46.20  &$<$43.33 &$<$43.50 & $<$46.03 & 46.41& $<$ 0.29\\
14 &  14503 & 3360 &45.59  &46.01  & $<$43.41& $<$43.59 & $<$46.11 & 45.92 &$<$ 0.60\\
15 &  14516 &719  &43.34  &44.00  & 42.14& 42.46 & 44.84 & 43.67 & $>$ 0.94 \\
16 &  14589  &1310  &45.01  & 45.09 & $<$42.55&$<$43.02 &$<$45.25 & 45.34 & $<$ 0.44 \\
17 &  15065 &1870  & 44.92 &45.31  & $<$42.54&43.34& 45.24 & $<$45.25 &$<$ 0.50 \\
18 &  15307  & 6030 & 46.06 &46.50  & $<$43.32&$<$44.09 & $<$46.02 & 46.39 & $<$ 0.30\\
19 &  15385 &2030  & 45.07 &45.59  &42.78 & $<$43.06& 45.48 & 45.40 & $>$ 0.55\\
20 &  15458 & 1270 &44.55  &44.89  & $<$42.05&$<$42.21 & $<$44.75 & 44.88 & $<$ 0.43\\
21 &  15492 & 1606 & 44.89 &45.14  & $<$42.43&$<$42.87 & $<$45.13 & 45.22 & $<$ 0.45\\
22 &  15496 &930 & 44.81 &45.18  & $<$42.26& $<$42.63 & $<$44.96 & 45.14 & $<$ 0.40\\
23 &  15585 & 2730 & 45.01 &45.15  & 43.25& 43.67& 45.95 & 45.34 & $>$ 0.80\\
24 & 16001 & 4050 & 46.17 &46.41  &$<$42.69 &$<$43.48 & $<$45.39 & 46.50 & $<$ 0.07\\
25 & 16073 & 1090 &44.78  &45.35  &43.03 & $<$43.22 & 45.73 & 45.11 & $>$ 0.80\\
26 &  16156 &620  &44.78  &45.09  & $<$42.11&$<$42.22 & $<$44.81 & 45.11 & $<$ 0.34\\
27 &  16242 &740  & 43.59 & 44.24 &$<$41.22 & $<$41.94 & $<$43.92 & 43.92 & $<$ 0.50\\
28 &  17233 &4270  & 45.07 &45.94  &$<$43.23 & $<$43.64 & $<$45.93 & 45.40 & $<$ 0.77 \\
\enddata
\tablenotetext{a}{Luminosity distance determined by E.L. Wright, http://www.astro.ucla.edu/~wright/CosmoCalc.html, for H$_0$ = 71 \kmsMpc, $\Omega_{M}$=0.27 and $\Omega_{\Lambda}$=0.73.}
\tablenotetext{b}{Total luminosity of starburst, assuming that log[L$_{ir}$(SB)] = log[L(6.2)]+ 2.7$\pm$0.1, from Brandl et al.(2006) and Pope et al.(2007).}
\tablenotetext{c}{Total luminosity from dusty torus of AGN, assuming log[L$_{ir}$(AGN)] = log[$\nu$L$_{\nu}$(5.5 \ums)] + 0.33$\pm$0.06, from Polletta et al.(2007) and assuming that all continuum luminosity $\nu$L$_{\nu}$(5.5 \ums) arises from an AGN.}
\tablenotetext{d}{Fraction of total luminosity which arises from a starburst.  If a PAH feature is detected, this fraction is shown as a lower limit because some of the continuum at 5.5\,\um which is attributed to an AGN actually arises from the starburst. If no PAH feature is detected, this fraction is shown as an upper limit because there is no direct evidence that a starburst is present. Source 9 is a blazar so the parameters for determining L$_{ir}$(AGN) and L$_{ir}$(SB) are not applicable.}

\end{deluxetable}

\clearpage
%
%
\begin{figure}
\figurenum{1}
\includegraphics[scale=1.7]{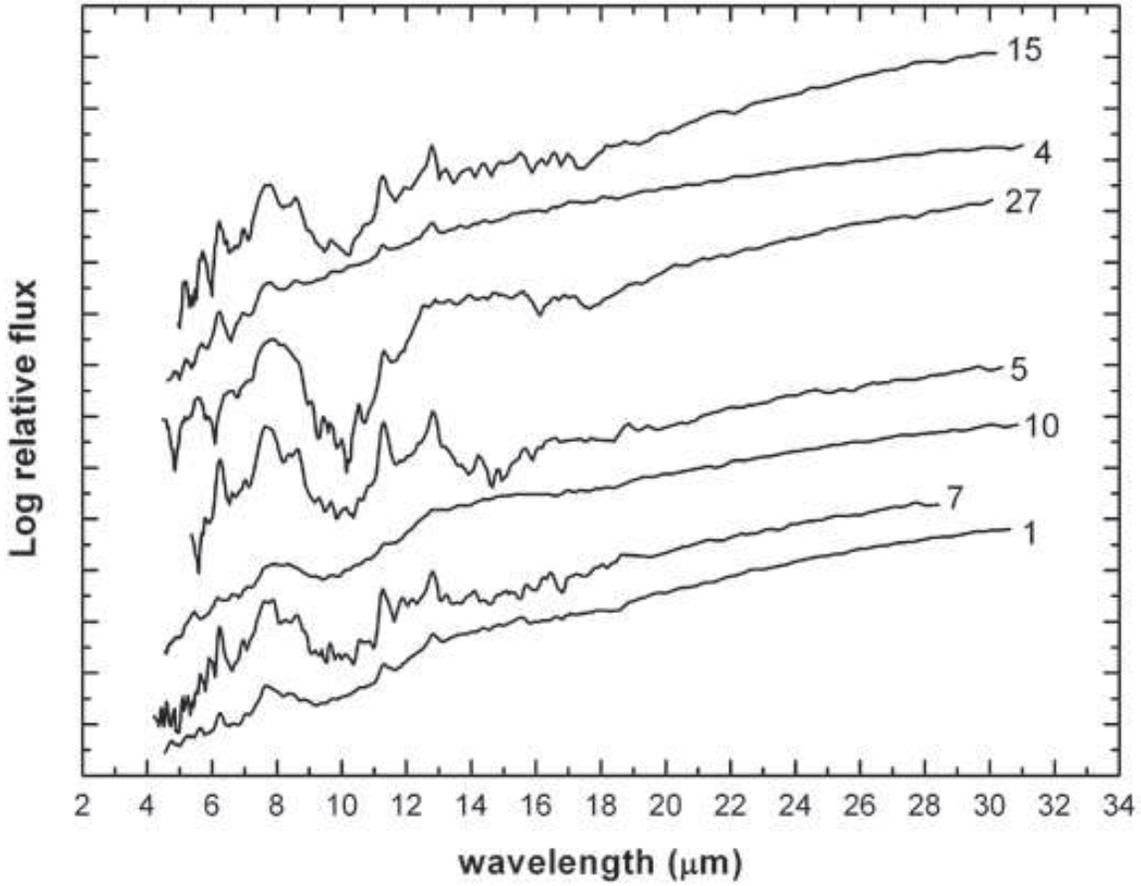}
\caption{IRS spectra of FSC sources in rest frame, ordered by $\nu$L$_{\nu}$(5.5 \ums) in Table 3, lowest luminosity at the top. Sources of higher luminosity continue in Figure 2.}

\end{figure}

\begin{figure}
\figurenum{2}
\includegraphics[scale=1.7]{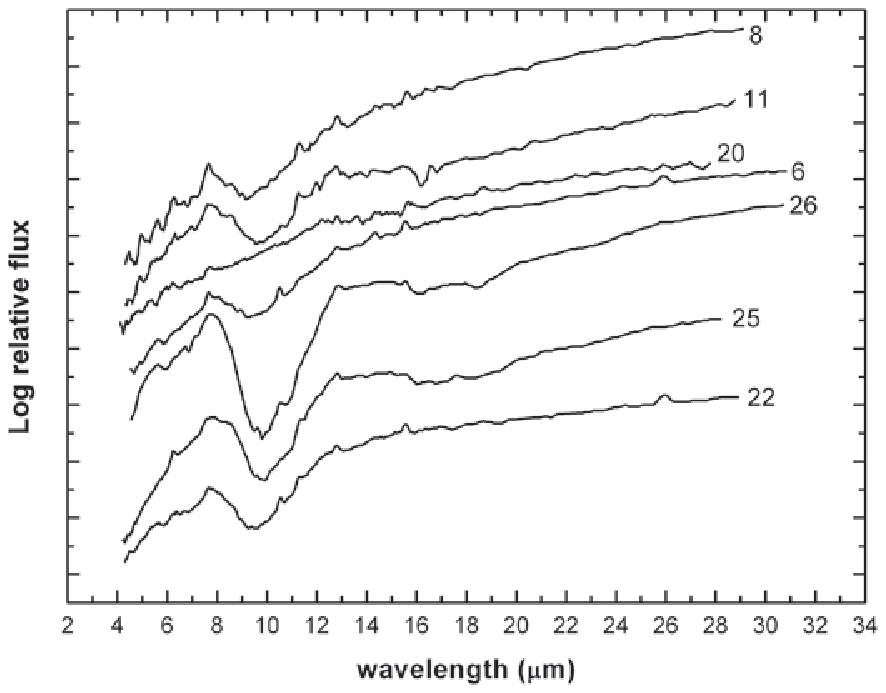}
\caption{IRS spectra of FSC sources in rest frame, ordered by $\nu$L$_{\nu}$(5.5 \ums) in Table 3, lowest luminosity at the top, continued from Figure 1. Sources of higher luminosity continue in Figure 3.}

\end{figure}

\begin{figure}
\figurenum{3}
\includegraphics[scale=1.7]{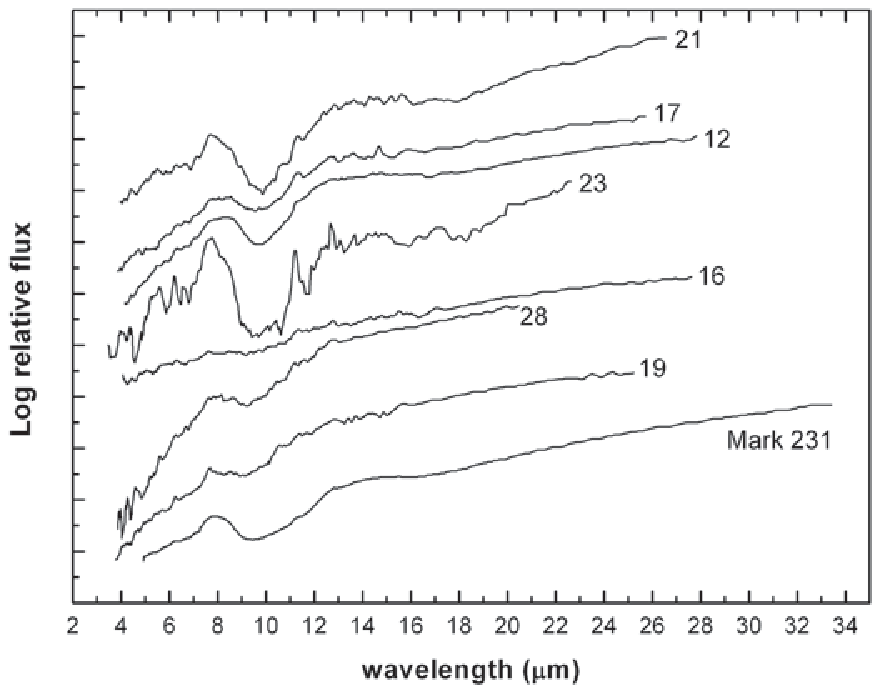}
\caption{IRS spectra of FSC sources in rest frame, ordered by $\nu$L$_{\nu}$(5.5 \ums) in Table 3, lowest luminosity at the top, continued from Figure 2. Sources of higher luminosity continue in Figure 4.}

\end{figure}

\begin{figure}
\figurenum{4}
\includegraphics[scale=1.7]{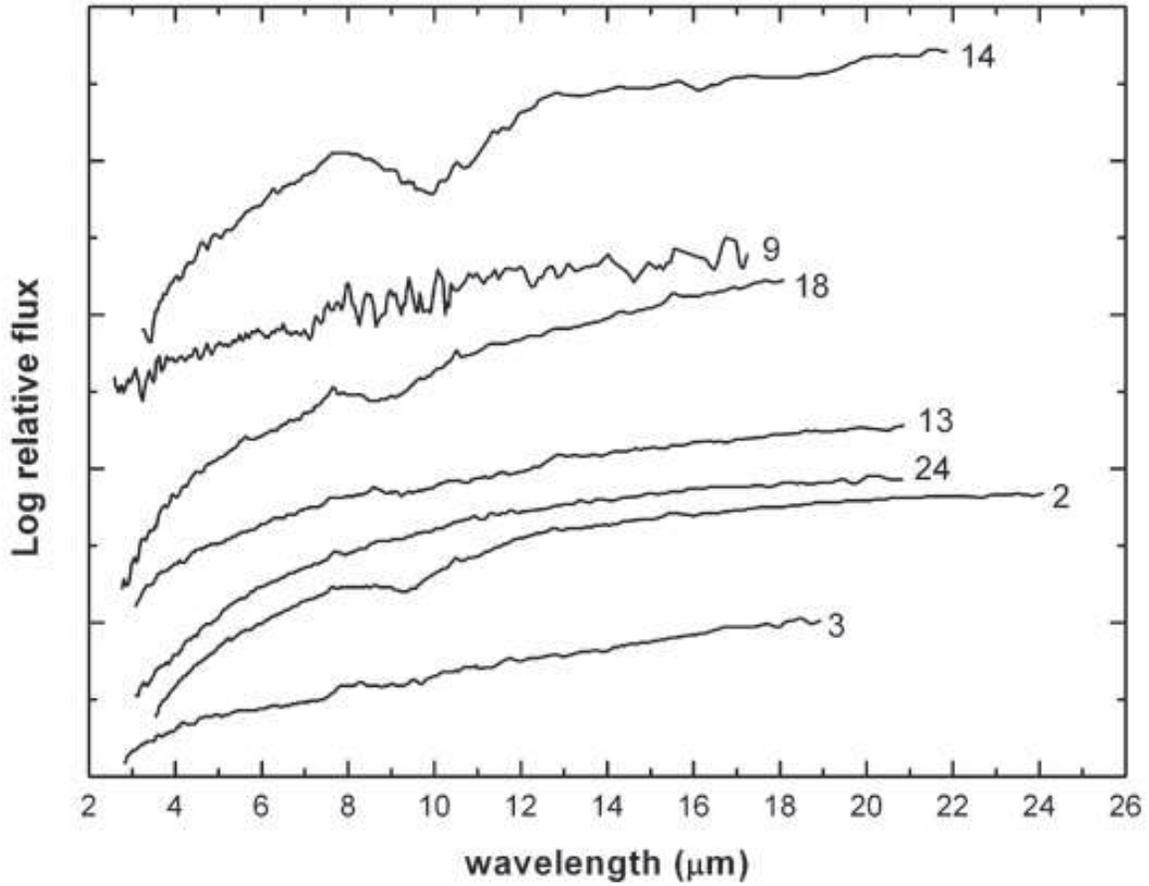}
\caption{IRS spectra of FSC sources in rest frame, ordered by $\nu$L$_{\nu}$(5.5 \ums) in Table 3, lowest luminosity at the top, continued from Figure 3. Source having highest luminosity is number 3.}

\end{figure}

\begin{figure}
\figurenum{5}
\includegraphics[scale=1.8]{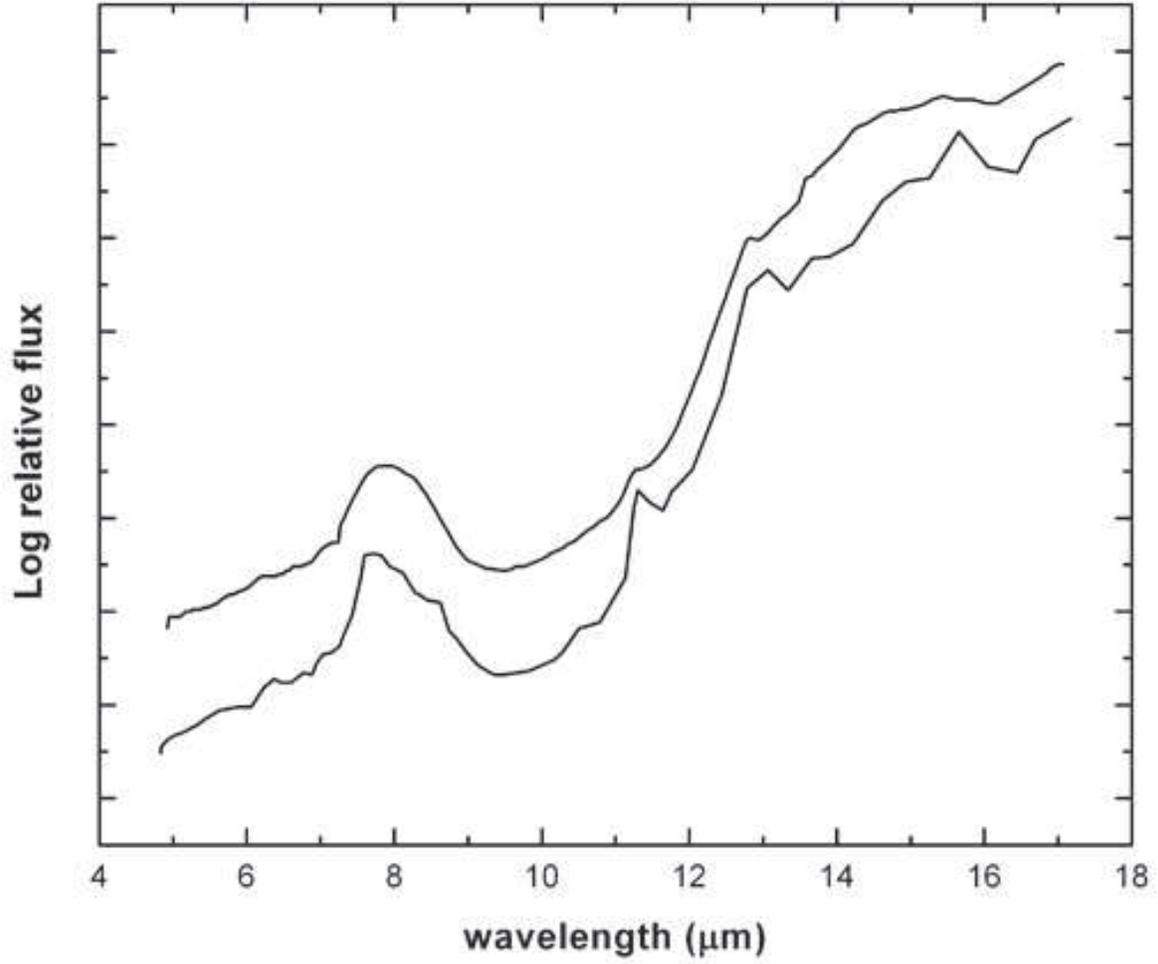}
\caption{Comparison of Markarian 231 (top) with average of the 19
FSC galaxies in Table 1 with measured silicate absorption (bottom),
arbitrarily normalized.}

\end{figure}

\begin{figure}
\figurenum{6}
\includegraphics[scale=1.4]{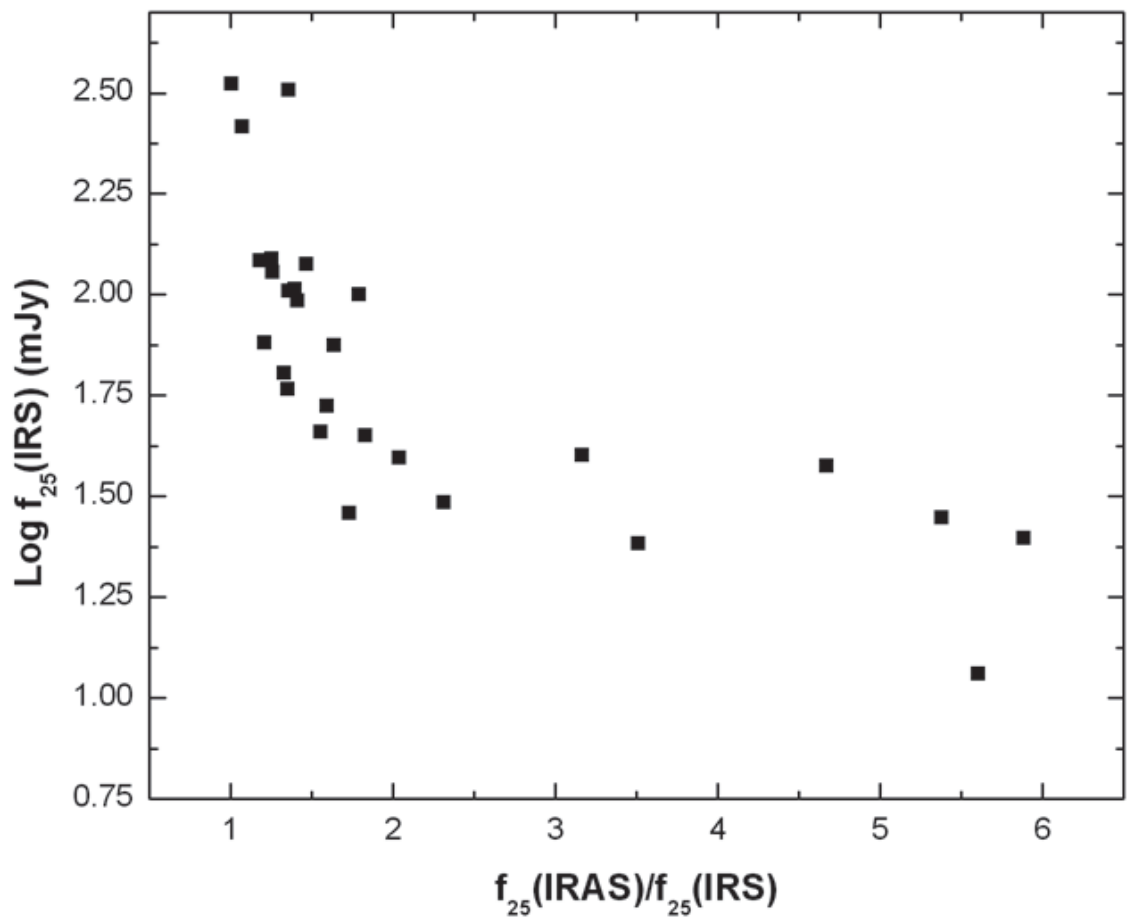}
\caption{Comparison of $f_{\nu}$(25 \ums) measured with $Spitzer$ IRS to $f_{\nu}$(25 \ums) listed in IRAS FSC for sources in Table 2. Uncertainties in IRS fluxes arise primarily from calibration uncertainties and are $\sim$ $\pm$ 5\%, comparable to the size of plotted symbols.  Large discrepancies between IRS and IRAS fluxes for fainter sources are discussed in text as indication that many IRAS sources are spurious.}

\end{figure}

\begin{figure}
\figurenum{7}
\includegraphics[scale=1.7]{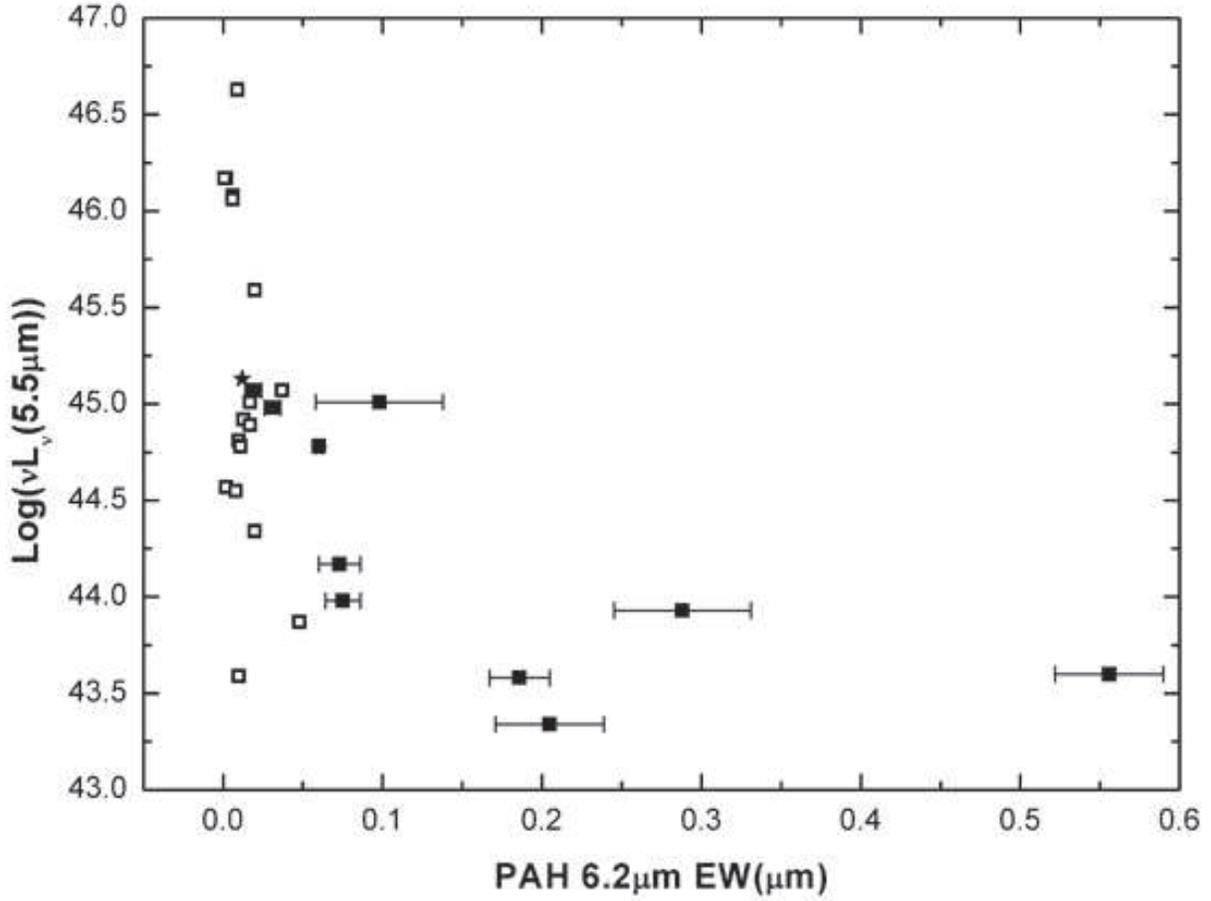}
\caption{Comparison of luminosities and PAH strengths for FSC sources; filled squares are sources with PAH detections and open squares show PAH upper limits for sources without PAH detections.  The star indicates Markarian 231. Error bars show uncertainties in PAH EWs from Table 2. }

\end{figure}

\begin{figure}
\figurenum{8}
\includegraphics[scale=1.4]{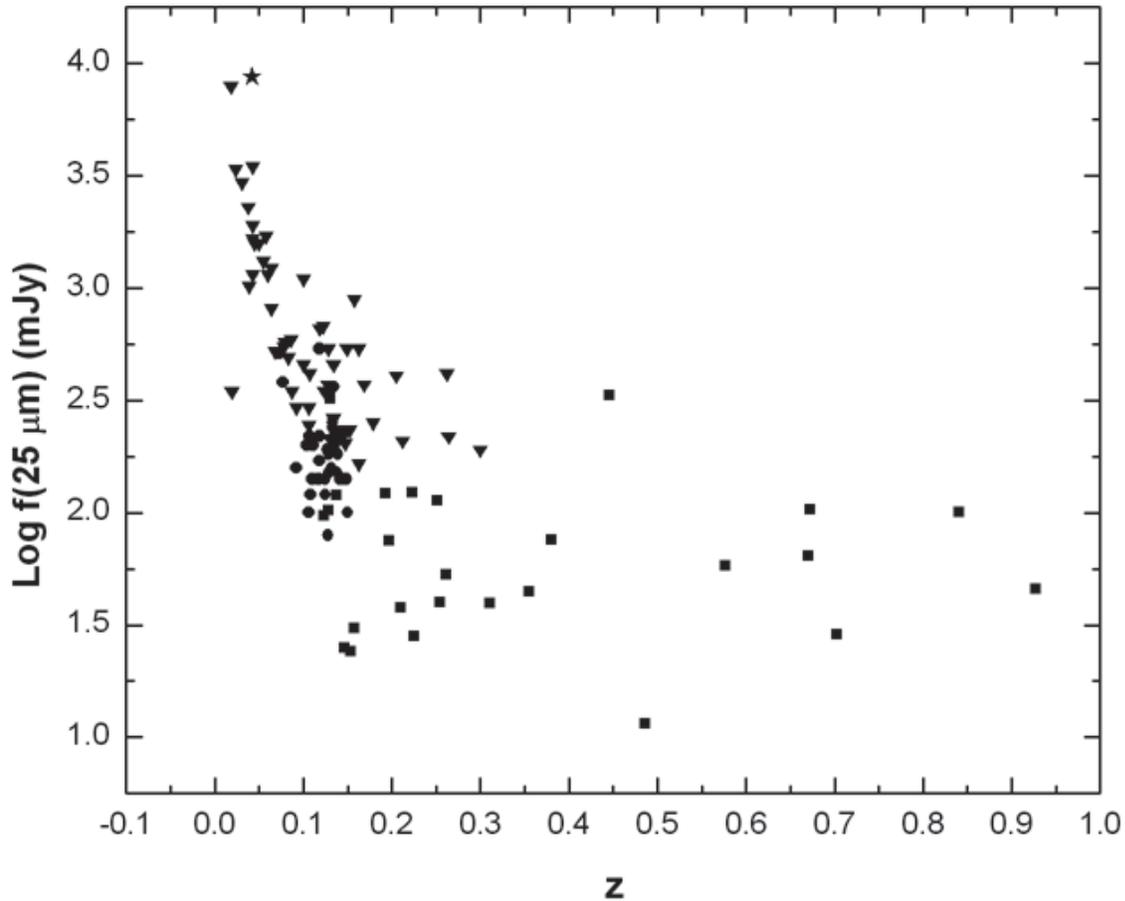}
\caption{Distribution of redshift and $f_{\nu}$(25 \ums) for ULIRG samples in Farrah et al. (2007)(filled triangles) and Imanishi et al. (2007) (filled circles) compared to FSC sample in present paper(filled squares). The star indicates Markarian 231. The $f_{\nu}$(25 \ums) shown for ULIRGS are taken from IRAS fluxes; the $f_{\nu}$(25 \ums) shown for FSC sources are synthetic $f_{\nu}$(25 \ums) that should be observed by IRAS as determined from $Spitzer$ IRS spectra. Uncertainties in IRS fluxes arise primarily from calibration uncertainties and are $\sim$ $\pm$ 5\%, comparable to the size of plotted symbols.}

\end{figure}

\begin{figure}
\figurenum{9}
\includegraphics[scale=1.3]{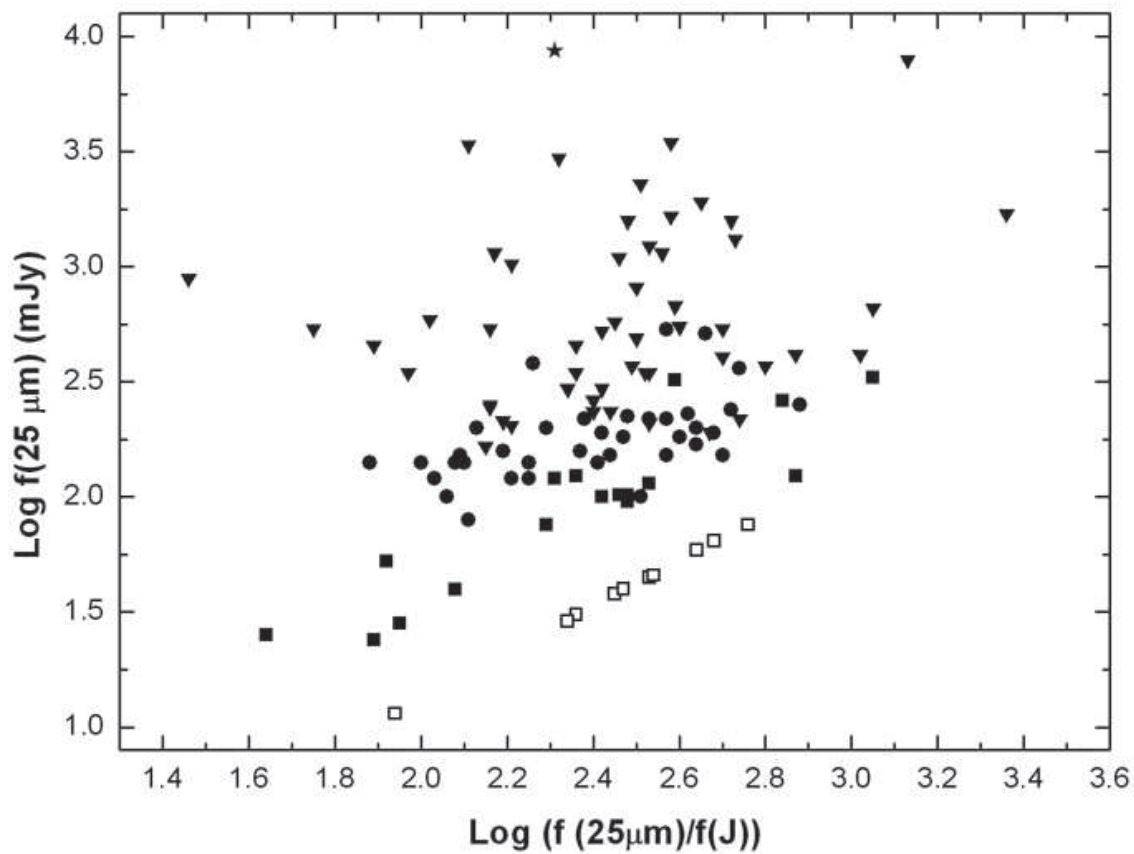}
\caption{Comparison of mid-infrared and near-infrared (2MASS J) fluxes for ULIRG samples in Imanishi et al. (filled circles), Farrah et al.(filled triangles), and  FSC sample in Table 2 (squares); open squares are limits for sources without J detections so the f(25 \ums)/f(J) ratio is greater than shown. The star indicates Markarian 231. The $f_{\nu}$(25 \ums) shown for ULIRGS are taken from IRAS fluxes; the $f_{\nu}$(25 \ums) shown for FSC sources are synthetic $f_{\nu}$(25 \ums) that should be observed by IRAS as determined from $Spitzer$ IRS spectra. Uncertainties in IRS fluxes arise primarily from calibration uncertainties and are $\sim$ $\pm$ 5\%, comparable to the size of plotted symbols.}

\end{figure}

\begin{figure}
\figurenum{10}
\includegraphics[scale=1.5]{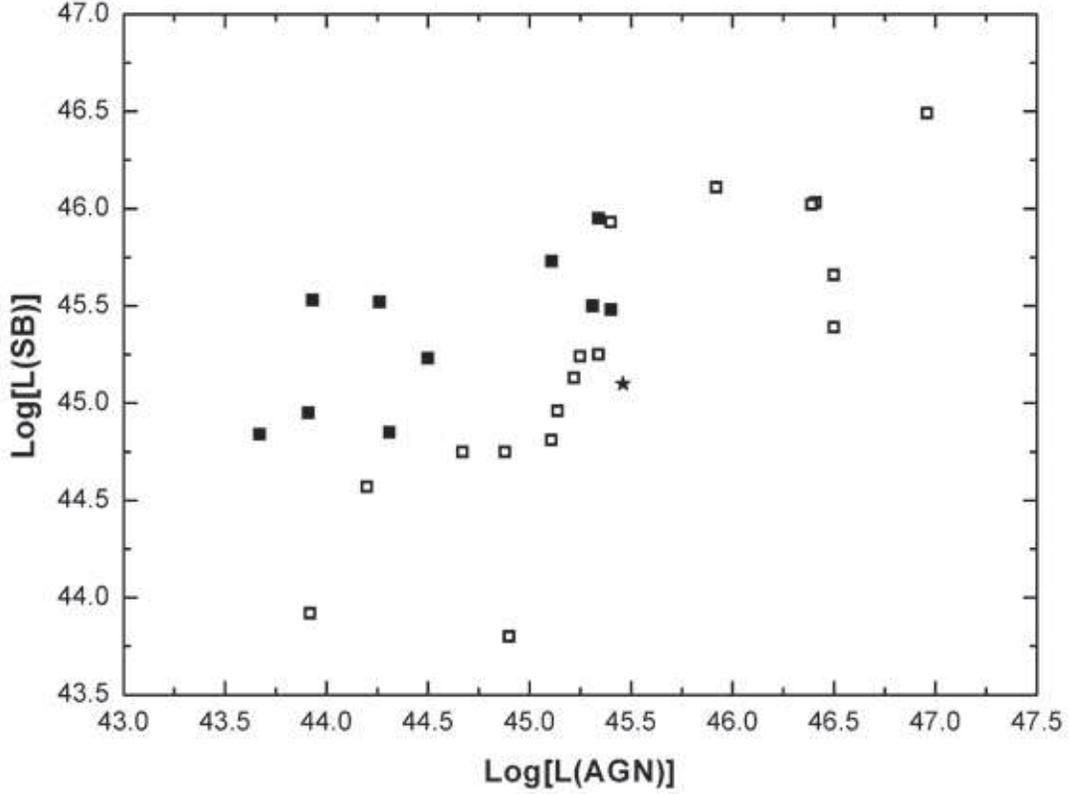}
\caption{Comparison of bolometric luminosities from AGN with those
from starbursts for FSC sample in Table 3, assuming that
log[L$_{ir}$(SB)] = log[L(6.2 \ums)]+ 2.7 for PAH luminosity L(6.2)
and that log[L$_{ir}$(AGN)] = log[$\nu$L$_{\nu}$(5.5 \ums)] + 0.33
for continuum luminosity $\nu$L$_{\nu}$(5.5 \ums). Filled squares
are starbursts with PAH detections so the L(AGN) derived from
continuum luminosities are upper limits for L(AGN) because some
continuum may arise from the starburst; open squares are sources
without PAH detections so the L(SB) derived from PAH luminosities
are upper limits for L(SB) determined from upper limits on L(6.2).
The star indicates Markarian 231.}

\end{figure}

\begin{figure}
\figurenum{11}
\includegraphics[scale=1.6]{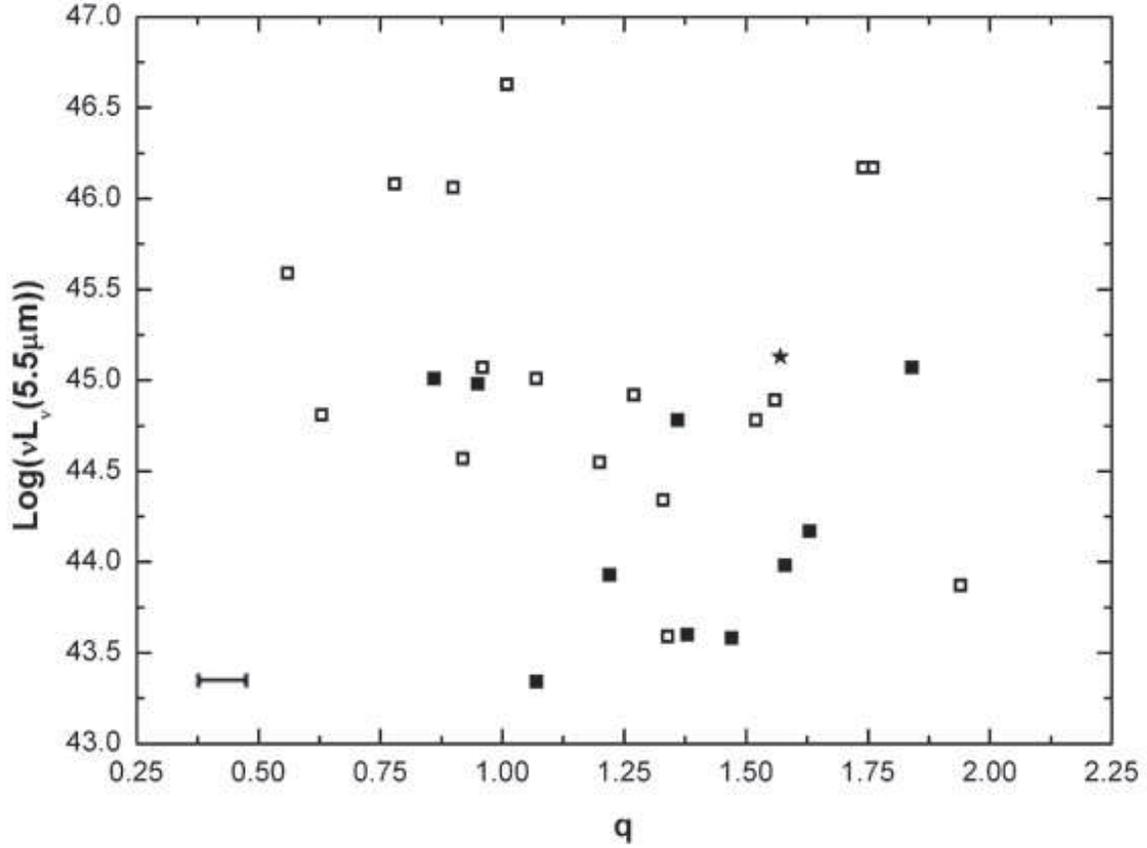}
\caption{Comparison of luminosities and radio strength for FSC sources; q = log[$f_{\nu}$(25 \ums)/$f_{\nu}$(1.4 GHz)] in observed frame.  Filled squares are starburst sources (as defined by PAH detections), and open squares are AGN (no PAH detections). The star indicates Markarian 231. Uncertainties in q are typically $\pm$ 10\%, shown by the representative error bar, arising primarily from uncertainty whether to adopt "peak" flux or "integrated" flux for the FIRST sources.}

\end{figure}

\end{document}